\documentclass[a4paper]{article}

\usepackage{spconf}
\usepackage{amsmath,graphicx}
\usepackage{amsfonts}
\usepackage{amsthm, color}
\usepackage{xspace}
\usepackage{multirow}
\usepackage{enumitem}
\usepackage{xcolor}
\usepackage{algorithm} 
\usepackage{algpseudocode} 
\usepackage{hyperref}
\usepackage{marvosym}

\usepackage{caption}
\usepackage{subcaption}

\hypersetup{
    colorlinks=true,
    citecolor=blue,
    linkcolor=blue,
    filecolor=magenta,      
    urlcolor=blue,
}

\newcommand{\aura}{\textit{Aura}\xspace}

\title{Aura: Privacy-preserving Augmentation to Improve Test Set Diversity in Speech Enhancement}
\name{ Xavier Gitiaux $^1$, Aditya Khant$^1$, Ebrahim Beyrami$^1$, Chandan Reddy \sthanks{Work performed while at Microsoft. Currently affiliated with Google.}$^2$, Jayant Gupchup\sthanks{Work performed while at Microsoft. Currently affiliated with Uber.}$^3$, Ross Cutler$^1$}
\address{
         1. Microsoft Corporation, Redmond WA, USA. Email: \emph{firstname.lastname@microsoft.com} \\
         2. Email: \emph{chandan.ka@outlook.com}, 3. Email: \emph{gupchup@gmail.com} \\
         }

\begin{document}

\maketitle
\begin{abstract}
Speech enhancement models running in production environments are commonly trained on publicly available data. This approach leads to regressions due to the lack of training/testing on representative customer data. Moreover, due to privacy reasons, developers cannot listen to customer content. This `ears-off' situation motivates  
\aura, an end-to-end solution to make existing speech enhancement train and test sets more challenging and diverse while being sample efficient. \aura is `ears-off' because it relies on a feature extractor and metrics of speech quality, DNSMOS P.835, and AECMOS, that are pre-trained on data obtained from public sources. We evaluate \aura on two speech enhancement tasks: noise suppression (NS) and audio echo cancellation (AEC). 

\aura samples an NS test set 0.42 harder in terms of P.835 OVRL than random sampling; and, an AEC test set 1.93 harder in AECMOS. Moreover, \aura increases diversity by 30\% for NS tasks and by 530\% for AEC tasks compared to greedy sampling. Moreover, \aura achieves a $26\%$ improvement in Spearman's rank correlation coefficient (SRCC) compared to random sampling when used to stack rank NS models.  

\end{abstract}

\section{INTRODUCTION}
Communication platforms such as Microsoft Teams and Skype operate under a wide range of devices, speakers, and ambient conditions, where speech enhancement (SE) tasks consist in removing background noise, reverberation, and echo \cite{ephraim1985speech}. Deep learning-based approaches provide state-of-the-art improvement of speech quality but typically involve supervised training on synthetic data that mix clean and noisy speech. 
Synthetic data cannot fully capture real-world conditions and test sets used during model evaluation

are not necessarily representative of all the audio scenarios present in

customer workloads \cite{xu2021deep}. 
Moreover, privacy and compliance often prohibit labelling customer workloads.

This paper presents \aura, an `ears-off' methodology to sample efficiently audio clips that are: (i) challenging to SE models and (ii) representative of the diverse conditions existing in the customer workload. In our setting, `ears-off' means that audio from customers cannot be listened to; only aggregate metrics can be logged out of the `ears-off' environment. 

\aura operates in an `ears-off' mode by relying on a pre-trained audio feature extractor such as VGGish \cite{hershey2017cnn} and pre-trained accurate objective speech quality metrics such as DNSMOS P.835 \cite{reddy2021DNSMOSP935} or AECMOS \cite{purin2022aecmos}. Both feature extractors and speech quality metrics are convolution-based models trained on large-scale open-sourced
data. 
\aura maps customer data into an embedding space, stratifies the data via clustering, identifies the most challenging scenarios for the SE task within each cluster, and reports aggregate model performances on these challenging conditions.

We apply \aura to rank $28$ noise suppression models from the INTERSPEECH 2021 DNS Challenge \cite{reddy2021interspeech} 
in terms of DNSMOS P.835 \cite{reddy2021DNSMOSP935}. We show that the SRCC between a $1\%$ sample produced by \aura and the entire mixture of noisy speech is $0.91$. This is an improvement of $26\%$ compared to a $1\%$ random sampling of noisy speech. We also apply \emph{Aura} to augment the benchmark test set for NS  
\cite{reddy2021interspeech}. The resulting test set is more challenging to NS models with a predicted absolute overall decline of $0.27$ in differential MOS and captures more diverse audio scenarios with an increase of $31\%$ in diversity (as measured by $\delta_{\chi^{2}}$ distance) compared to the current benchmark test set \cite{reddy2021interspeech}. Lastly, we use \aura to create a test set for AEC models with an absolute decline of 1.92 in differential echo MOS relative to random sampling and a $530\%$ increase in diversity relative to greedy sampling.

The main contributions of this paper are as follows: a `ears-off' 
methodology to construct challenging and diverse 
test sets for SE applications deployed in production environments; a generic approach to measure test set diversity; 
    \footnote{Code is available at \href{https://github.com/microsoft/aura}{github.com/microsoft/Aura}.}.


\section{Related Work}
\label{sec:relatedwork}
The Deep Noise Suppression (DNS) benchmark challenge \cite{reddy2021interspeech} created an extensive open-sourced noisy speech test set. 
However, the resulting benchmark test set does not cover all the scenarios experienced by customers using real-time communication platforms. \aura addresses this gap by creating a test set more representative of customer workload while preserving customer privacy.

The literature on test set construction in production scenarios is sparse. In software development, Rothermel et al.\ present metrics to evaluate testing coverage \cite{rothermel1996analyzing}: path coverage, percentage of tests affecting system state, and worst-case runtime metrics. Mani et al.\ evaluate the coverage of test sets for classifiers using class distribution (equivalence partitioning), within-class distance (centroid positioning), and between class statistics (boundary conditioning) \cite{mani2019coverage}. 
\aura extends Mani et al.\ to speech enhancement problems where there are no predictive classes to measure testing coverage. Instead,  \aura uses classes 
from a known ontology of audio events \cite{45857} and replaces centroid positioning and boundary conditioning statistics with a speech quality metric that quantifies the level of difficulty in the overall test set. 

Well-established ontologies are instrumental in creating large-scale and diverse public dataset in object detection (e.g., ImageNet, \cite{deng2009imagenet}, Open Images \cite{kuznetsova2020open}) and audio event detection (Audio Set \cite{45857}). 
ImageNet relies on the hierarchical ontology established by Wordnet \cite{miller1995wordnet} and contains about 15 million images that cover 22K Wordnet concepts. 
Audio Set relies on 
an ontology of 632 sound types. Pham et al.\ show how to measure diversity based on an existing ontology, a reference distribution and a distance metric \cite{pham2010visualization}. 

Active learning is commonly employed to prioritise the samples to label when labelling is expensive \cite{settles2009active,kossen2021active}. The learner selects the data points from a pool of unlabelled data using either uncertainty-based or diversity-based heuristics. Kossen et al.\ show that it is optimal to prioritise hard examples to reduce the variance of a given evaluation metric \cite{kossen2021active}. However, these uncertainty-based heuristics do not preclude selecting redundant scenarios. Sener et al.\ propose a diversity-based solution identifying the most representative set of data points that are the most representative of the entire data set \cite{sener2017active}. 

 This paper integrates diversity and uncertainty-based techniques to create a test set for SE models.

A growing body of work has developed differentially private learning algorithms that do not expose private customer data \cite{abadi2016deep}. Differential privacy allows private computation of evaluation metrics \cite{boyd2015differential} if the data is already annotated. 
Federated learning is a partial solution to the `ears-off' problem as it enables training models on decentralised devices 
\cite{latif2020federated}. However, federated learning does not solve the issue of collecting an `ears-off' test set to evaluate the model on a production workload. Finally, we use DNSMOS P.835 \cite{reddy2021DNSMOSP935} and AECMOS \cite{purin2022aecmos} to estimate speech quality without the need for a clean reference and for listening to the audio content.

\section{Use Cases and Solution}
\label{sec:soln}

In this section, we present the use cases for \aura 
and describe our end-to-end solution to address each use case.

\subsection{Test set 1 (TS1): Challenging and Diverse Test Set}
\label{subsec:testset}
We apply \aura first to sample a subset of audio clips from a database 
to form for a given SE model a test set with 
challenging and diverse scenarios. The quality of the test set is measured 
with an objective speech quality metric ($\rho$) and a diversity metric ($\delta_{\chi^{2}}$) (see Section \ref{subsec:e2e}).

\subsection{Test set 2 (TS2): Production Workload Test Set}
\label{subsec:ranking}
A typical model evaluation task is to rank the speech quality $\rho$ produced by multiple SE models against a representative workload that 
mirrors the distribution observed in the production system. The main motivation is to 
(i) ensure that SE models do not introduce regressions by degrading the scenarios frequently encountered by users of the system,
and (ii) find 
audio scenarios that allow discriminating model performances. We rank SE models according to the quality metric $\rho$. We measure sampling performance by comparing the Spearman's rank correlation coefficient (SRCC) between the ranking of the models obtained with the sample and the ranking obtained with the entire representative workload.

\begin{figure}
    \centering
    \includegraphics[width=0.99\columnwidth]{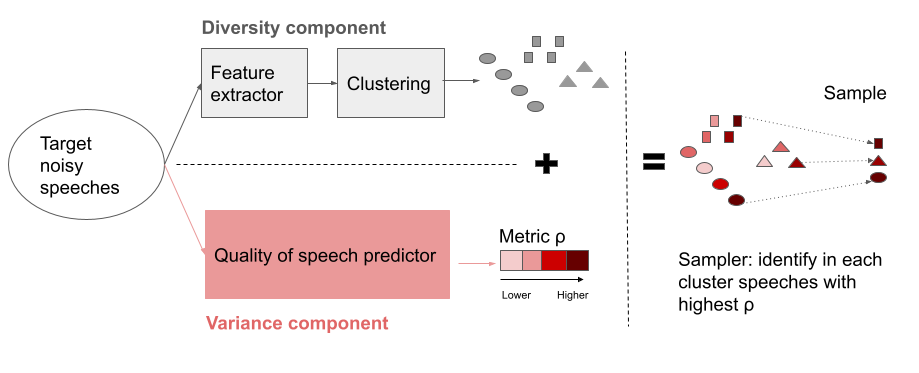}
    \caption{Overall structure of \emph{Aura}. \emph{Aura} reduces the variance of the performance metric $\rho$ (red) while maximising the diversity component (gray) to ensure good coverage of audio scenarios in the embedding space. 
    }
    \label{fig:structure}
\end{figure}

 \begin{algorithm}[t]
	\caption{Diversity evaluation} 
	\begin{algorithmic}[1]
	\State \textbf{Inputs}: Sample $\mathcal{S}$, noise type ontology $\mathcal{O}$
	\State \textbf{Output:}  $\chi^{2}$ distance $\delta_{\chi^{2}}$
	\State $n=|S|$: \# of clips in $\mathcal{S}$
	\State $C=|\mathcal{O}|$: \# of classes in $\mathcal{O}$
    \State  $p_{c}=\frac{\mbox{number of clips in } \mathcal{S} \mbox{ with noise type } c}{n}$   
	\State $p_{u}=\frac{1}{C}$ (probability of classes in uniform distribution)
		\State Compute $\delta_{\chi^{2}}=\displaystyle\sum_{c=1}^{C}\frac{\left(p_{c}- p_{u}\right)^{2}}{p_{c}+ p_{u}}$
		\State Returns  $\delta_{\chi^{2}}$.
	\end{algorithmic} 
	\label{alg:diversity}
\end{algorithm}

\subsection{End-to-end Solution}
\label{subsec:e2e}
The end-to-end \aura system is described in Figure \ref{fig:structure}. 

\noindent
\textbf{Input:} A large database of speech clips ($\mathcal{D}$). For TS1 (Section \ref{subsec:testset}), the clean speech clips are filtered out. For TS2 (Section \ref{subsec:ranking}), the database contains the representative workload that includes clean speech.

\noindent
\textbf{Output:} A test set sampled from the input dataset.

\noindent
\textbf{Feature Extractor:} Our extractor is constructed from VGGish, a model pre-trained on a 100M YouTube videos dataset \cite{hershey2017cnn}. It generates a 128-dimension embedding for each clip. 

\noindent
\textbf{Clustering:} In practice, audio clips collected in a production environment are not labelled with sound types from the Audioset ontology. Instead, we create a soft ontology of sound types by performing kmeans++ on the embedding space to partition $1.5$ million noisy speech into $256$ clusters \cite{arthur2006k}. The number of clusters is selected as the number that achieves the lowest Davies-Bouldin index, where the index is computed using Euclidean distance in the embedding space.

\noindent
\textbf{Sampler:} To trade-off variance and bias, \aura 
applies probability-proportional-to-size sampling within each cluster. Within each cluster, it prioritises clips with the highest $\rho$. 

\noindent
\textbf{Objective Metric:} For each speech clip, the P.835 \cite{naderi_subjective_2021} protocol generates a MOS for signal quality SIG, background noise BAK, and overall quality OVRL; and the P.831 protocol \cite{IC3-AI-P831} evaluates echo impairment. Reddy et al.~and Purin et al.~show that convolution-based models DNSMOS and AECMOS achieve 0.98 correlation with the human ratings obtained via the P.835 and P.831 listening protocols \cite{reddy2021DNSMOSP935,purin2022aecmos}. We derive our quality of speech metric $\rho$ from DNSMOS and AECMOS by measuring the differential MOS (DMOS) between DNSMOS and AECMOS predictions before and after denoising. A negative value for the DMOS $\rho$  indicates that the SE method degrades the quality of speech.

\section{EVALUATION AND RESULTS}
\label{sec:results}
\subsection{Diversity Metric} 
We leverage the ontology from Audio Set and follow the approach presented by Pham et al.\ \cite{pham2010visualization}. We 
consider a uniform distribution over the 527 sound types in Audio Set as our reference distribution. We measure the diversity of a test set as the 
histogram distance $\delta_{\chi^{2}}$ \cite{pele2010quadratic} between the reference distribution and the distribution of the test set (see Algorithm \ref{alg:diversity})  
The lower the $\chi^{2}$ distance, the more audio properties encoded in the embedding space the resulting test set covers: $\delta_{\chi^{2}}=0$ means that the test set covers uniformly all noise types; $\delta_{\chi^{2}}=1$ means that the test set has no overlap with the noise categories in the reference distribution.
\subsection{Simulated Ears-off Dataset}
We perform experiments on a simulated ears-off dataset that serves as as a ground truth and is made publicly available \cite{auragithub}.
For the NS task, we create noisy speech candidates to add to the benchmark test set \cite{reddy2021interspeech}. We mix sounds from the balanced split of Audio Set \cite{45857} with clean speech clips \cite{reddy2021interspeech}. We use segments from Audio Set as background noise and follow \cite{reddy2021interspeech} to generate noisy speech with a signal-to-noise ratio (SNR) between -5 dB and 5 dB. The resulting target data combines 1000 files from the benchmark test set with 22K of 10 seconds clips from the newly synthesised noisy speech that covers 302 noise types. 
The pool of noisy speech candidates is partially out-of-distribution compared to \cite{reddy2021interspeech}, which covers 227 noise types, and 
the train set of 
DNSMOS P.835 and thus, reproduces the conditions of an `ears-off' environment.
 
For each noisy speech in the previous target data, we randomly draw 10 clean speech clips from \cite{reddy2021interspeech}. The clean speech presents a challenge

to stack rank models because it does not allow discriminating the performance of noise suppressors. For the AEC task, we collected 33K audio clips from the INTERSPEECH 2021 AEC Challenge dataset that contains diverse speaker and noise conditions

\cite{cutler2021interspeech}.

\setlength{\belowcaptionskip}{-5pt}
\begin{table}[]
    \centering
     \caption{\aura-based DNS test set.  Sample provided by \aura on the augmented benchmark dataset increases difficulty level (lower DMOS) and diversity (lower $\delta_{\chi^{2}}$, higher coverage)). 
     $\pm$ represents 95\% conf. intervals and ``classes covered'' represents the categories covered in the Audio Set ontology. 
     } 
    \begin{tabular}{|c|ccc|c|c|}
    \hline
    Test set & \multicolumn{3}{c|}{DMOS} & $\delta_{\chi^{2}}$ & Classes \\
        &  SIG & BAK & OVRL &  & covered\\
           \hline
    DNS & -0.24 & 0.99 & 0.18 & 0.53 & 227\\
    Benchmark \cite{reddy2021interspeech}& \scriptsize{$\pm 0.01$} & \scriptsize{$\pm 0.01$} & \scriptsize{$\pm 0.01$} & &\\
    \hline 
   
    Random   & -0.23 & 1.27 & 0.33 & 0.55  & 243 \\
        & \scriptsize{$\pm$  0.01} & \scriptsize{$\pm 0.01$} & \scriptsize{$\pm$  0.01} & & \\
    
    \hline
       Greedy   & -0.57& 0.91 &-0.32 & 0.53 &  249 \\
       (no cluster) & \scriptsize{$\pm$  0.01} &  \scriptsize{$\pm 0.01$} & \scriptsize{$\pm$  0.01}&   &\\
       \hline 
       \aura  & -0.40 & 0.92 & -0.09 &0.37 & 302  \\
        & \scriptsize{$\pm 0.01$} & \scriptsize{$\pm 0.01$} & \scriptsize{$\pm 0.01$} &  &\\
     \hline
    \end{tabular}
    
    \label{tab:sampling}
\end{table}

\noindent

\subsection{TS1: Challenging and Diverse Test Set}
\label{subsec:ts1}
\textbf{Noise suppression.} For generating TS1, we only consider the noisy speech candidates from the simulated ears-off dataset and create a test set of 1000 clips as in \cite{reddy2021interspeech}. 
We measure the DMOS metric $\rho$ and $\delta_{\chi^{2}}$ as explained in Section \ref{subsec:e2e}. We say that a test set is challenging if it captures the audio scenarios for which $\rho \leq 0$. 

Table \ref{tab:sampling} compares DMOS and $\delta_{\chi^{2}}$ for the following sampling strategies: random, greedy (sorted by ascending DMOS without clustering), and \aura. The \aura test set is more challenging to NS models than the benchmark and random test sets. NS models degrade by $0.16$ in SIG DMOS and $0.27$ in OVRL DMOS compared to the benchmark test set. 
Covering each cluster in the embedding space is instrumental to increasing the diversity of the test set. \aura's test set with clustering covers $302$ noise types in the reference distribution (out of $527$ classes), while \aura's test set without clustering (Greedy) covers only $249$ of the noise types, which is only a marginal improvement over the benchmark test set. \aura balances diversity and challenging scenarios, while greedy sampling could oversample one challenging but redundant scenario.
 
The $\chi^{2}$ distance decreases from $0.53$ without clustering to $0.37$ with clustering. 

\begin{figure}[]
    \centering
    \includegraphics[width=0.45\columnwidth]{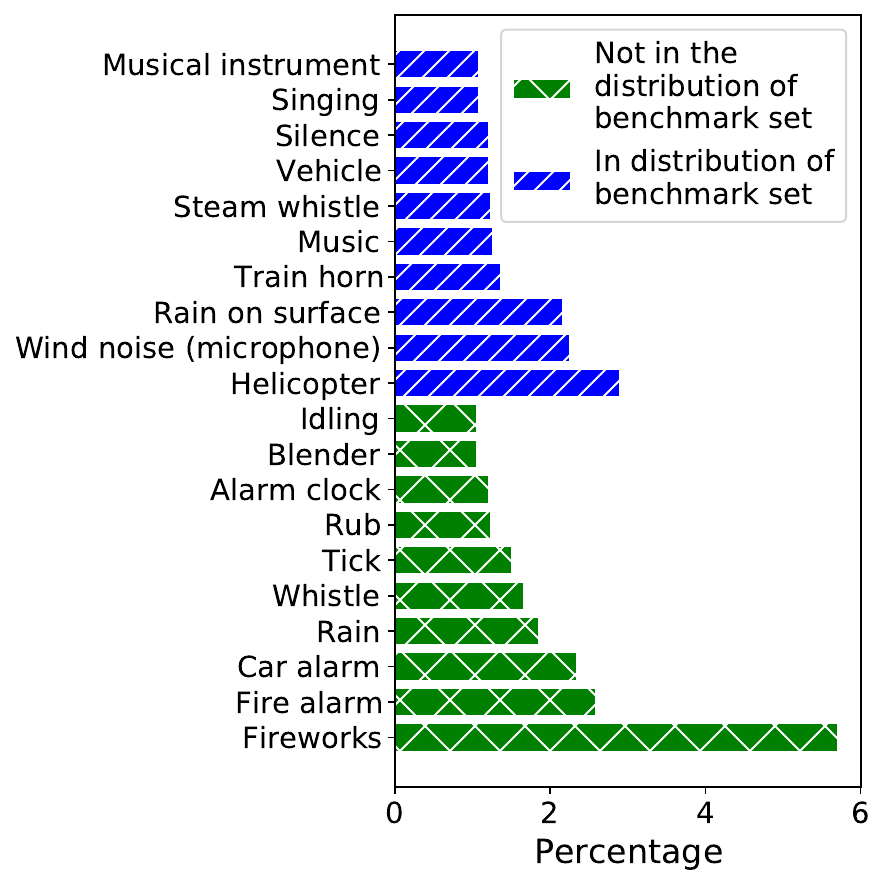}
    \caption{Top-10 categories not in distribution (green) and in-distribution (blue) noise categories in \aura test set. In-distribution relates to categories in the benchmark test set  \cite{reddy2021interspeech}.}
    \label{fig:Top-10}
\end{figure}

Figure \ref{fig:Top-10} shows the top-10 noise types (in green) from the augmenting dataset of noisy speech that \emph{Aura} adds to the test set. \emph{Aura} prioritises new audio scenarios 
compared to the top 10 noise types present in the benchmark test set.

\textbf{Audio Echo Cancellation.}
Table \ref{tab:sampling-aec} shows similar results for echo cancellation tasks. The \aura test set is significantly more challenging than random with an absolute decline in echo MOS of 1.93 relative to random. \aura reduces the DMOS gap between tests created by random sampling and the most challenging test set (as produced by greedy sampling) that can be obtained from our simulated dataset by 84\%. Unlike greedy sampling, \aura's decrease in DMOS comes with a significant increase in the diversity of audio conditions included in the test set: the $\delta_{\chi^{2}}$ decreases from 0.76 (for Greedy) to 0.12 for \aura. 
\setlength{\belowcaptionskip}{-5pt}
\begin{table}[]
    \centering
     \caption{\aura-based AEC test set.  Same as in Table \ref{tab:sampling} but for AEC task.
     } 
    \begin{tabular}{|c|c|c|c|}
    \hline
    Test set & DMOS & $\delta_{\chi^{2}}$ & Classes \\
        &   &  & covered\\
           \hline
    AEC  Benchmark& 2.78 \scriptsize{$\pm0.01$} & 0.58 & 75\\
    \hline 
   
    Random   & 1.74 \scriptsize{$\pm 0.01$ } & 0.59  & 88 \\
    \hline
       Greedy (no cluster)  & -0.55\scriptsize{$\pm 0.01$} & 0.76 & 79 \\
       \hline 
       \aura  & -0.19\scriptsize{$\pm 0.01$} & 0.12 & 147\\
     \hline
    \end{tabular}
    
    \label{tab:sampling-aec}
\end{table}

\textbf{Sensitivity to the number of clusters.}
\aura identifies the optimal number (256) of clusters to seed kmeans++ by minimising the Davies-Bouldin index (see section \ref{subsec:e2e}). In Figure \ref{fig:cluster_sigmos} (left), OVRL, SIG, and BAK DMOS are not sensitive to increases in the number of clusters from 64 to 512. For signal quality (SIG), the DMOS remains stable around -0.4. 

\subsection{TS2: Production Workload Test Set}
\label{subsec:ts2}
\begin{table}[]
\caption{SRCC between the ranking of 28 NS models obtained from a $1\%$ \aura sample with the ranking obtained from using the entire data. 95\% conf interval depicted using $\pm$.}
    \centering
    \resizebox{\columnwidth}{!}{\begin{tabular}{|c|ccc|}
    \hline
         Sampling & \multicolumn{3}{c|}{SRCC}  \\
         method &SIG&BAK&OVRL \\
         \hline
         \emph{Random} & $0.58$\scriptsize{$\pm 0.02$} & $0.72$\scriptsize{$\pm 0.01$}& $0.72$ \scriptsize{$\pm 0.02$}  \\
         \hline
        \emph{Stratified-Random} & $0.72$\scriptsize{$\pm 0.02$} & $0.89$ \scriptsize{$\pm 0.01$} & $0.82$\scriptsize{$\pm 0.01$}  \\
         \hline
         \emph{Variance} & $0.80$\scriptsize{$\pm 0.01$} & $0.93$ \scriptsize{$\pm 0.01$} & $0.88$\scriptsize{$\pm 0.01$}  \\
         \hline
         \emph{Aura} & \textbf{0.84}\scriptsize{\textbf{$\pm$ 0.01}} & \textbf{0.93} \scriptsize{\textbf{$\pm$ 0.01}}& \textbf{0.91}\scriptsize{\textbf{$\pm$ 0.01}} \\
          \hline
    \end{tabular}}
    \label{tab:spearman}
\end{table}

\begin{figure}
\centering
\begin{minipage}{.25\textwidth}
  \centering
  \includegraphics[width=.75\linewidth]{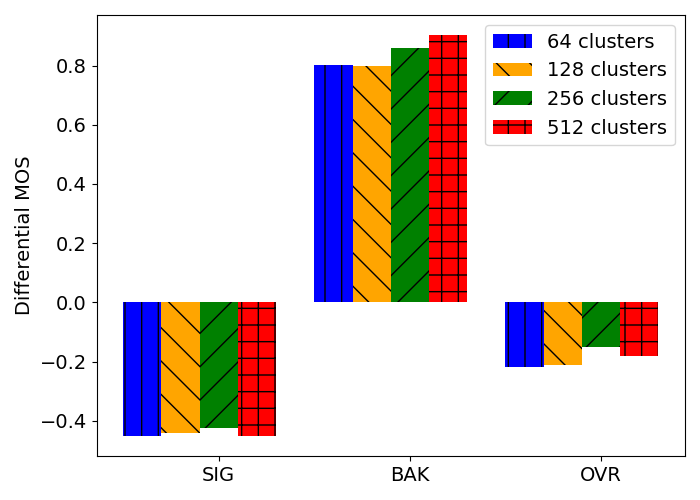}
\end{minipage}%
\begin{minipage}{.25\textwidth}
  \centering
  \includegraphics[width=.75\linewidth]{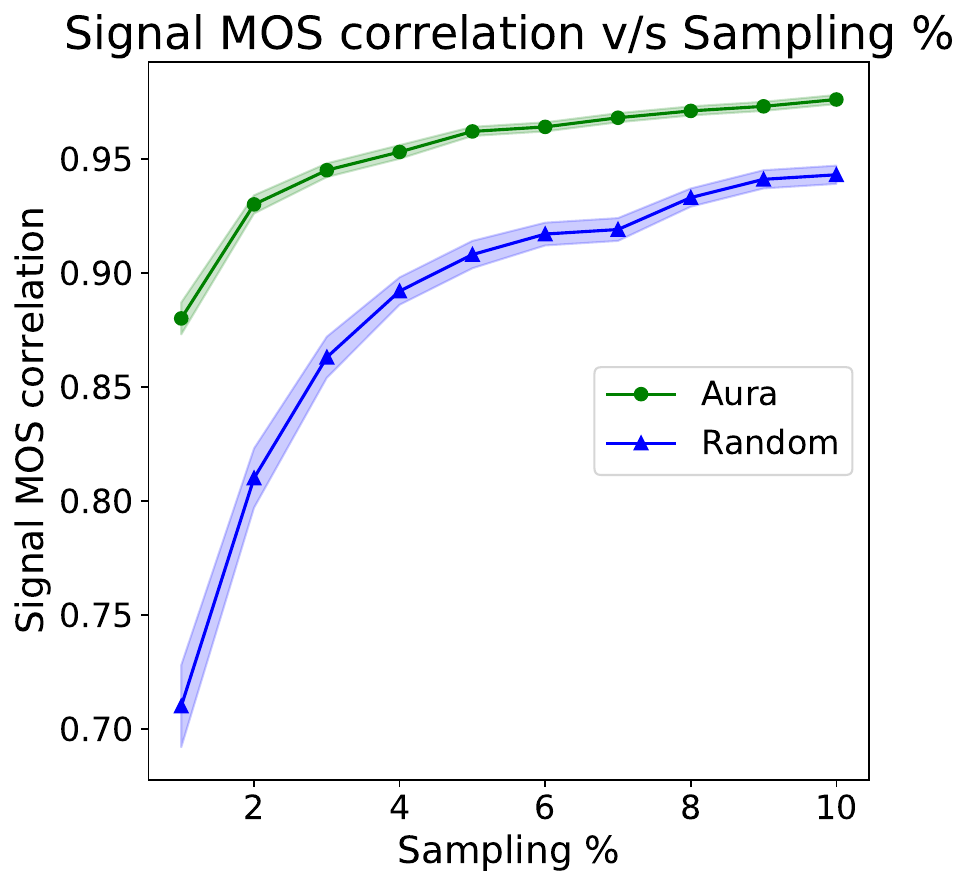}
\end{minipage}
\caption{Left: \aura's sensitivity to the number of clusters. This shows DMOS for NS task and \aura-based test sets generated with different numbers of clusters. Right: SRCC of \aura-based rank estimates as a function of sample size.}
\label{fig:cluster_sigmos}
\end{figure}

For TS2, we include clean speech in the simulated dataset. For each speech clip, we run $28$ noise suppression models from the benchmark challenge \cite{reddy2021interspeech}. We report the SRCC of the DNSMOS P.835 for the sample compared to the entire dataset. We bootstrap the sampling 200 times and report the mean and standard deviation of the resulting rank correlation coefficients. We compare \emph{Aura}'s sampling performance to three alternatives: (i) \emph{Random}, which draws randomly $1\%$ of data; (ii) \emph{Stratified-Random}, which stratifies the data into clusters and samples uniformly within clusters; (iii) \emph{Variance}, which samples proportionally to the variance of DMOS across the 28 models. In Table \ref{tab:spearman}, 
\emph{Aura}'s sampling leads to an improvement in SRCC across all three P.835 components. For overall quality (OVRL), we observe a $26\%$ SRCC improvement over random sampling. We also observe that the $95\%$ confidence interval of the ranking obtained from \emph{Aura}-based samples are narrower than the one obtained by random sampling.
Figure \ref{fig:cluster_sigmos} (right) shows that for larger samples, \emph{Aura} still outperforms \emph{Random} in terms of rank correlation.

\section{Conclusion}
\label{sec:conclude}
\aura presents an end-to-end system for improving test sets used to evaluate speech enhancement models in an `ears-off' environment.
\aura creates challenging, diverse test sets by relying on objective quality metrics, a pre-trained feature extractor, and an established ontology. The method is generic and can be applied to diverse audio tasks, including noise suppression and echo cancellation.

\aura balances customer privacy and needs to measure model performance in real-world scenarios.

Future work includes (i) adding new classes that appear in customer workload and are not yet covered by the existing sound type ontology; and, (ii) sampling train set for SE models.

\pagebreak
\bibliographystyle{IEEEtran}

\bibliography{references,IC3-AI}

\end{document}